\documentclass[12pt, preprint]{aastex}
\usepackage{txfonts}
\usepackage{longtable}
\usepackage{natbib}
\renewcommand{\cite}{\citealp}

\shorttitle{On the impact of helium content on the RR Lyrae distance scale}
\shortauthors{Marconi et al.}

\begin{document}

\title{On the impact of helium content on the  RR Lyrae distance scale}

 \author{M. Marconi\altaffilmark{1}, 
G. Bono\altaffilmark{2,3}, 
A. Pietrinferni\altaffilmark{4}, 
V. F. Braga\altaffilmark{5,6}, 
M. Castellani\altaffilmark{3}, 
R.F. Stellingwerf\altaffilmark{7}
}

\altaffiltext{1}{INAF-Osservatorio astronomico di Capodimonte, Via
  Moiariello 16, 80131 Napoli, Italy; marcella.marconi@oacn.inaf.it}
\altaffiltext{2}{Dipartimento di Fisica - Universit\`a di Roma Tor Vergata, Via della Ricerca Scientifica 1, 00133 Roma; giuseppe.bono@roma2.infn.it}
\altaffiltext{3}{INAF-Osservatorio Astronomico di Roma, Via Frascati 33, 00078 Monte Porzio Catone, Italy; marco.castellani@oa-roma.inaf.it}
\altaffiltext{4}{INAF-Osservatorio Astronomico d'Abruzzo, Via M. Maggini SNC, 64100, Teramo, Italy; adriano@oa-teramo.inaf.it}
\altaffiltext{5}{Instituto Milenio de Astrofisica, Santiago, Chile; vittorio.braga@roma2.infn.it}
\altaffiltext{6}{Departamento de Fisica, Facultad de Ciencias Exactas, Universidad Andres Bello, Fernandez Concha 700, Las Condes, Santiago, Chile}
\altaffiltext{7}{Stellingwerf Consulting, 11033 Mathis Mtn Rd SE, 35803 Huntsville, AL USA; rfs@swcp.com}

 \begin{abstract}

We constructed new sets of He-enhanced (Y=0.30, Y=0.40) nonlinear,
time-dependent convective hydrodynamical models of RR Lyrae (RRL) stars
covering a broad range in metal abundances (Z=0.0001--0.02). The increase 
in He content from the canonical value (Y=0.245) to Y=0.30--0.40 causes a
simultaneous increase in stellar luminosity and in pulsation period.
To investigate the dependence of the RR Lyrae distance scale on the
He abundance we computed new optical ($RI$) and near-infrared ($JHK$) 
Period-Luminosity-Metallicity-Helium (PLZY) relations. Interestingly
enough, the increase in He content causes a minimal change in the coefficients
of both period and metallicity terms, since canonical and He--enhanced models
obey similar PLZ relations. On the contrary, 
the classical $B$- and $V$-band mean magnitude-metallicity
relations and the R-band PLZ relation display a significant dependence 
on the He content. The He-enhanced models are, at fixed metal content, 
0.2-0.5 mag brighter than canonical ones. This variation is only 
marginally affected by evolutionary effects. 
The quoted distance diagnostics once calibrated with trigonometric parallaxes 
({\it Gaia}) will provide the opportunity to estimate the He content of 
field and cluster RRLs. Moreover, the use of either spectroscopic or 
photometric metal abundances will pave the way to new empirical constraints 
on the universality of the helium--to--metal enrichment ratio in old 
(t$\gtrsim$ 10 Gyr) stellar tracers.
\end{abstract}



\section{Introduction} 

During the last century RR Lyrae (RRL) stars have played a crucial role as 
standard candles and tracers of old stellar populations 
~\citep{marconi15, madore17, neeley17}. 
They are old (t$\gtrsim$10 Gyr) low--mass radial variables in their central 
helium burning phase and are observed in the Milky Way 
~\citep{vivaszinn2006,zinn14,drake13,pietrukowicz2015}, in Local Group 
~\citep{soszynski10b,fiorentino12a,coppola15} and in 
Local Volume galaxies ~\citep[][]{dacosta10,sarajedini12}. 

RRLs are used as standard candles since they obey a relation between 
absolute visual magnitude and iron 
abundance~\citep[][]{caputo00,cacciari03a,dicriscienzo04}.
This relation, whose linearity has also been questioned in the 
literature \citep{caputo00,catelan04,dicriscienzo04}, suffers from
significant intrinsic errors and systematics.
RRLs do not obey a Period-Luminosity (PL) relation in the
optical bands but, thanks to the characteristic behaviour of
near-infrared (NIR) bolometric corrections~\citep{bono01,bono03c}, 
they obey a PL relation in the NIR 
regime~\citep[][]{longmore1990,braga15,coppola15}.

The advantages of these relations are the small dependence on
reddening and evolutionary effects~\citep[][]{bono03c} and a milder 
dependence on metallicity when compared with $B$,$V$ magnitudes.
Theory and observations indicate that more metal--rich RRLs
are fainter than metal-poor ones, but we still lack firm constraints 
on the coefficient of the metallicity term in the NIR PL relations ~\citep[][]
{bono03c,catelan04,dallora04,sollima06a,marconi15}. 

Optical and  NIR Period-Wesenheit (PW) relations are solid diagnostics
to determine individual RRL distances, but rely on the assumed reddening 
law~\citep[][]{dicriscienzo04,braga15,coppola15,marconi15}.
These relations are reddening free by construction 
~\citep[][]{madore82,ripepi12,riess2012,fiorentino13,inno13} and include a color term. 
This means that they mimic a Period-Luminosity-Color (PLC)
relation, tracing the position of each variable inside the instability strip (IS). 
These are the reasons why Period-Wesenheit relations have been 
widely adopted to trace the 3D structure of highly reddened clusters 
in the Galactic Bulge ~\citep{soszynski14,pietrukowicz2015}.

The main motivations for the current investigations are the following. 

a) The helium-to-metals enrichment ratio ($\Delta\,Y$/$\Delta\,Z$=1.4, with a 
primordial He abundance of 0.245) adopted in evolutionary 
\citep[][]{pietrinferni2006} and pulsation \citep[][]{marconi15} calculations 
is still affected by large uncertainties. RRLs are good laboratories for 
estimating the He content \citep{caputo1998}. To provide a new spin on the 
determination of this parameter we are investigating new pulsation observables 
together with spectroscopic measurements of the metal content for field and 
cluster RRLs.\par    
b) Using the $\Delta$S method \citet{walker1991a} found that 
Bulge RRLs approach solar metallicity. This finding  was recently supported 
by \citet{chadid2017} using high-resolution spectra, since they found several 
RRLs at solar chemical compositions. This means a metallicity regime in which 
RRL pulsation properties are more prone to helium effects \citep{bono95a,marconi11}. \par   
c) The RRL distance scale is going to play a crucial role to constrain possible 
systematics affecting primary distance indicators \citep{beaton2016}. Sizable 
samples of RRLs have already been identified in Local Group galaxies 
\citep{monelli2017} and beyond \citep{dacosta10}. However, we still lack 
firm theoretical and empirical constraints on the $\Delta\,Y$/$\Delta\,Z$ 
ratio in extragalactic systems. \par
To overcome the limitations of the current theoretical framework we computed 
new sets of pulsation models with the same metal abundances (Z=0.0001--0.02)
adopted in \citet{marconi15} but helium enriched (Y=0.30 and 
Y=0.40\footnote{Metals (Z) and helium (Y) abundances by mass fraction.},
Marconi et al. 2018, in preparation).

\section{Impact of helium-enhanced models on RRL distances}

Following the same prescriptions as in \citet{marconi15} for both
evolutionary and pulsation computations, and the same seven  metal abundances, 
new sets of He-enhanced RRL models were computed with Y=0.30 and Y=0.40.
The entire set of HB models ~\citep[][]{pietrinferni2006} are available 
in the BaSTI data base\footnote{http://www.oa-teramo.inaf.it/BASTI/}.
They were computed, for each assumed chemical composition, using a fixed 
core mass and envelope chemical profile and evolving a progenitor from 
the pre-main sequence to the tip of the Red Giant Branch with an age 
of $\sim$13 Gyr. For each chemical composition, the mass distribution 
of HB models ranges from the mass of the progenitors (coolest HB models) 
down to a total mass of the order of 0.5 $M_{\odot}$ (hottest HB models).

The evolutionary phases off the Zero-Age-Horizontal Branch (ZAHB)
have been extended either to the onset of thermal pulses, for more
massive models, or until the luminosity of the model (along the
white dwarf cooling sequence) becomes fainter than 
$\log (L/L_\odot)\sim$-2.5 for less massive structures.
The $\alpha$-elements were enhanced with respect to 
the~\citet{grevesse1993}\footnote{Note that new solar 
abundances by \citet{asplund2009} provide lower CNO 
abundances when compared with \citet{grevesse1993}. The difference in metal 
distributions mainly causes a difference in the zero-point. The impact on the 
HB mass-luminosity relation is within the different luminosity 
levels adopted, at fixed mass and chemical composition, to construct 
pulsation models.}
solar metal distribution by variable factors (see Table~1 
in~\citet{pietrinferni2006}). The overall enhancement---[$\alpha$/Fe]---is 
equal to 0.4 dex.

\begin{figure*}[t]
\centering
\includegraphics[width=18cm]{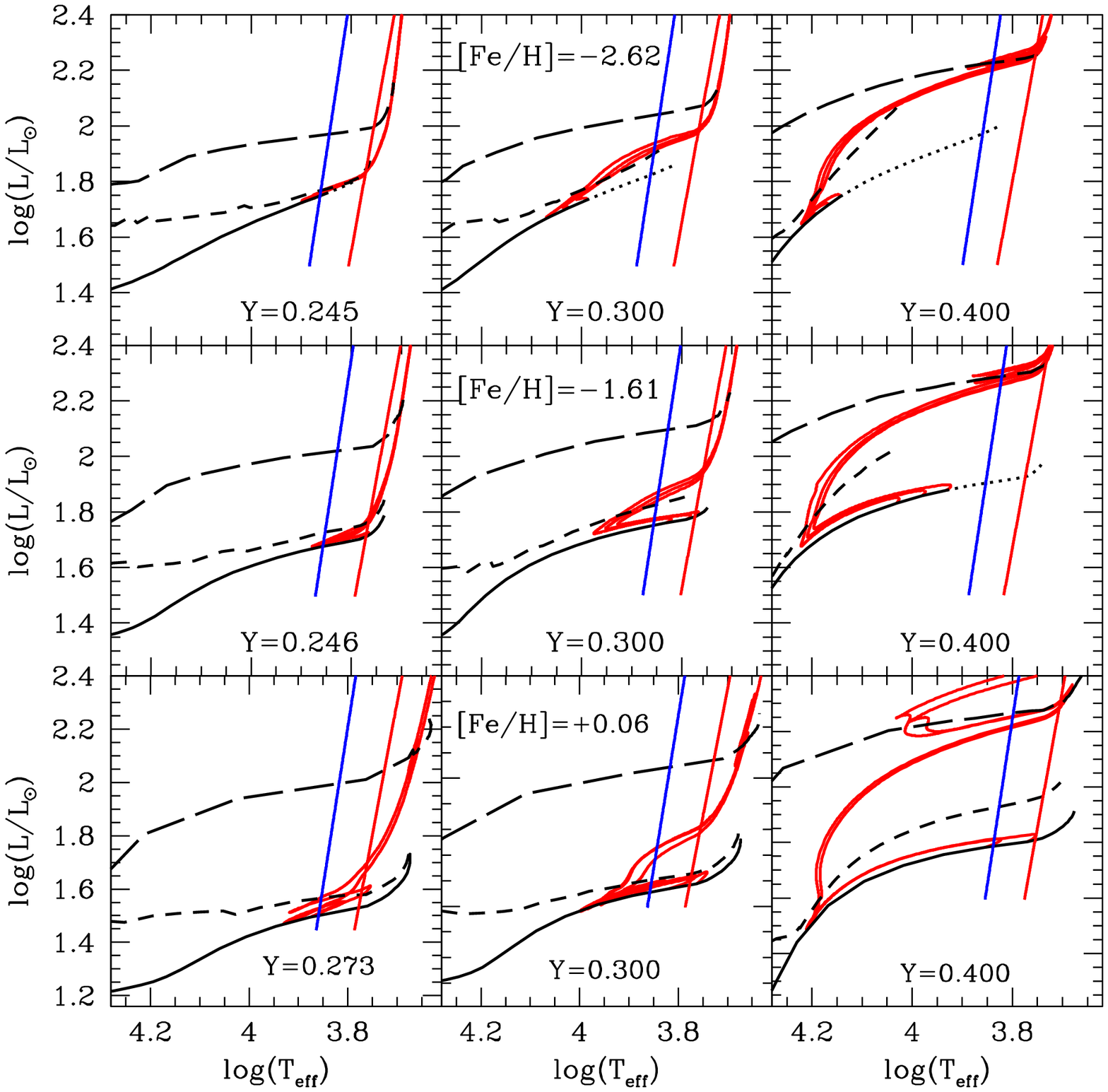}
\vspace*{-2.0truecm}
\caption{Hertzsprung-Russell diagrams of low-mass helium burning stellar structures.
Top -- from left to right evolutionary prescriptions for a metal-poor iron abundance
(see labeled value) and three different helium contents. The solid lines display the
Zero-Age-Horizontal-Branch (ZAHB), while the dotted lines show the extension to younger
progenitors till the crossing of the RRL IS. The dashed and the
long--dashed lines display the 90\% and the 100\% central helium exhaustion.
The red solid lines display three HB evolutionary models covering 
the mass range typical of RRLs located close to the blue edge, 
to the middle of the IS and to the red edge of the IS.
The blue and the red, almost vertical, lines show the blue and the red edge of the
predicted RRL IS.
Middle -- Same as the top, but for a metal-intermediate chemical composition.
Bottom -- Same as the top, but for a more metal-rich chemical composition.
}
\label{fig:teo1}
\end{figure*}

Fig.~\ref{fig:teo1} shows the behaviour of HB models in the Hertzsprung-Russell 
diagram for three assumptions on the helium and on the metal content. 
In each panel the black solid line shows the
location of the ZAHB, the dashed black line corresponds to a central helium
exhaustion at $90 \%$ level and the long dashed line to the complete
exhaustion. Note that the ZAHB becomes dotted for masses
higher than the progenitor one, artificially included to populate the
IS. The blue and the red vertical lines display the predicted 
blue and red edge of the IS. The red solid lines show
selected  evolutionary models of HB structures populating the RRL IS. 
They range from 0.76 $M_{\odot}$ to 0.80$M_{\odot}$ 
for Z=0.0001, Y=0.245  (top left panel), 
and from 0.520 $M_\odot$ to 0.525 $M_\odot$ for Z=0.0164, Y=0.400 
(bottom right panel).
Evolutionary prescriptions plotted in Fig.~\ref{fig:teo1} bring forward some relevant 
properties concerning He-enhanced stellar structures worth being discussed.

{\em i)}  {\em HB morphology}--He-enhanced stellar 
populations, at fixed metal content 
and cluster age, are characterized by smaller stellar masses at the main 
sequence turn off. This means smaller stellar masses at the tip of the 
red giant branch and a HB morphology dominated by hot 
and extreme stars. The HB luminosity function is, therefore, dominated 
by stars that are hotter than the blue edge of the IS.    
These stellar systems can still produce RRLs, since hot HB stars cross 
the IS just before or soon after the AGB phase 
\citep[post early AGB,][]{greggio90,dcruz1996}. This means that 
the red HB and the IS are 
poorly populated (see middle and right panels of Fig.~\ref{fig:teo1}).  

{\em ii) Evolutionary time scale inside the instability strip}--
The evolutionary time spent by a canonical, metal-poor (Z=0.0001, Y=0.245)
stellar structure (M=$0.84 M_\odot$) inside the IS during 
central He burning phases is $t_{HB}\sim$67 Myr. This time decreases 
by at least a factor of two when moving to He-enhanced models with Y=0.30 
(M=$0.82 M_\odot$, $t_{HB}\sim$32 Myr) and by a factor of six for models with 
Y=0.40 (M=$0.82 M_\odot$, $t_{HB}\sim$11 Myr). The quoted trend marginally  
changes with the metal content,  and indeed, canonical models at solar iron 
abundance (Z=0.0198, Y=0.273, M=$0.5450 M_\odot$) spend inside the IS 
an evolutionary time of $t_{HB}\sim$29 Myr and this time decreases down to 
$t_{HB}\sim$15 Myr for Y-0.30 (M=$0.5425 M_\odot$) and to $t_{HB}\sim$5 Myr 
for Y=0.40 (M=$0.5230 M_\odot$). The consequence of this difference is that 
the number of RRLs produced by He-enhanced stellar populations is at least 
one order of magnitude smaller than canonical ones. The reader is referred 
to Marconi et al. (2018) for more details.         

{\em iii)}  {\em Evolutionary time-scale to approach the instability strip}--The 
increase in helium content causes, at fixed metallicity, a steady decrease in 
the evolutionary time scale required to approach the IS. 
This effect is more severe in the metal-poor regime 
(see top panels in Fig.~\ref{fig:teo1}) 
where the ZAHB for He-enhanced models located inside the IS  
is populated by stellar structures that are significantly younger than typical RRLs. 
Canonical stellar structures with Z=0.0001 and Y=0.245 evolves from the 
Zero Age Main Sequence (ZAMS) to the ZAHB portion located inside the 
IS on a time scale of $\sim$12.5 Gyr \citep{valcarce2012}.
A He-enhanced stellar structure with Z=0.0001 and Y=0.30 evolves from 
the ZAMS to the IS in $\sim$8.5 Gyr, while for Y=0.40 the 
same time scale becomes of the order of $\sim$4.5 Gyr. A similar trend is 
also present among stellar structures with Z=0.0006 and Y=0.40, since 
they approach the portion of the ZAHB located inside the IS  
with ages younger than $\sim$8.5 Gyr.
On the other hand, for Z=0.001 and Y=0.40 the ZAHB stellar
structures located inside the IS have ages that are only
marginally younger (11--12 Gyr) than canonical ones.  Note that the
current empirical evidence indicates that RRLs stars have {\it only}
been identified in stellar populations older than 10 Gyrs 
\citep{dekany2018}.

On the basis of the quoted evolutionary prescriptions, we computed a set of 
pulsation models, for each iron and helium abundance, accounting for two 
values of the stellar mass and three different luminosity levels. 
The reasons for this choice were already discussed in \citet{marconi15}. 
Here we only give the highlights: 
1) the ZAHB mass and luminosity level as based on the adopted evolutionary models; 
2) the ZAHB mass and a luminosity level 0.1 dex brighter than the ZAHB luminosity; 
3) a stellar mass 10$\%$ smaller than the ZAHB value and a luminosity level 
0.2 dex brighter than the ZAHB luminosity.

The different sets of models were constructed following the same approach 
discussed in \citet{marconi15}. The bolometric light curves were 
transformed into  optical ($UBVRI$) and NIR
($JHK$)\footnote{We adopted the 2MASS---$JHK_s$---photometric system.}
bands using static atmosphere models \citep{bono95b} and eventually 
intensity weighted mean magnitudes and colors were computed. 
Preliminary results for interpreting Galactic Bulge RRLs were discussed 
in \citet{marconiminniti2018}, while the details of these new helium 
enriched models are presented in Marconi et al. (2018, in preparation).

\subsection{Predicted optical/NIR PL relations}

\begin{figure}
\includegraphics[scale=0.9]{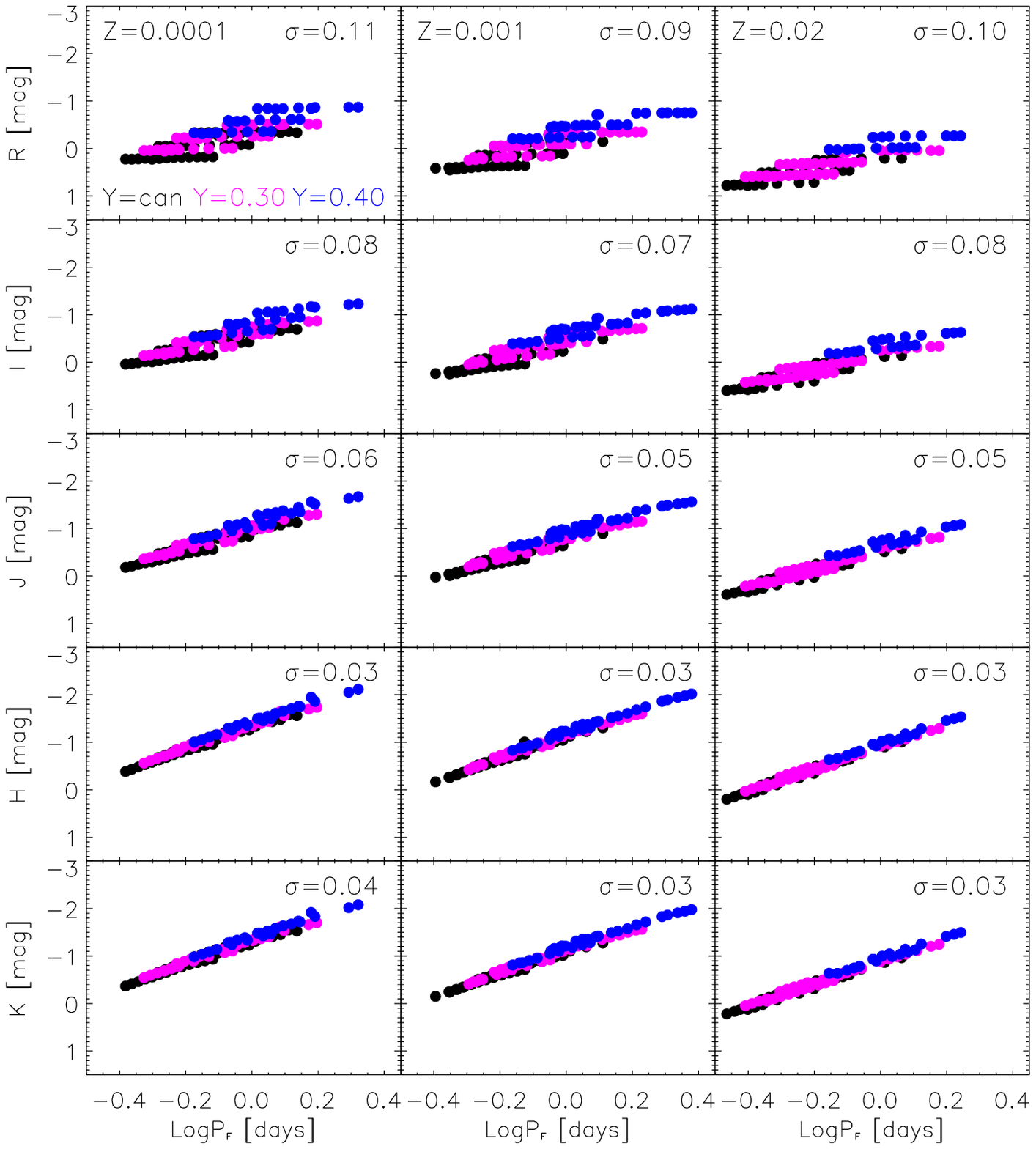}
\caption{From top to bottom predicted global (fundamental plus first overtone) 
PL relations in the $R$,$I$,$J$,$H$,$K$ bands for three different metal abundances: 
Z=0.0001 (left), Z=0.001 (central) and Z=0.02 (right).
Models plotted in each panel take account for different helium abundances: 
canonical Y as in \citet{marconi15} (black), Y=0.30  (magenta) and Y=0.40 (blue).
The standard deviations of the PL relations are labeled in the top right corner.}
\label{FIG1}
\end{figure}

Figure~\ref{FIG1} shows the predicted optical/NIR PL distribution 
for five bands ($R$,$I$,$J$,$H$,$K$) and for three different metal abundances: 
Z=0.0001 (left), Z=0.001 (middle) and Z=0.02 (right).
In each panel are plotted pulsation models constructed assuming a fixed 
helium-to-metal enrichment ratio \citet[][black circles]{marconi15}
together with models constructed assuming two different helium 
enhancements: Y=0.30  (magenta circles) and Y=0.40 (blue circles).
Models plotted in this figure display two well defined trends among 
canonical and helium enhanced models. 
{\em i)} The period distribution of helium enhanced models is systematically 
shifted towards longer periods when compared with canonical models. 
The difference is mainly caused by an evolutionary effect: a decrease 
in the mean stellar mass populating the RRL IS and 
an increase in the luminosity level \citep{marconi11}.  
{\em ii)} The spread in luminosity between canonical and helium enhanced 
models steadily decreases when moving from the $I$- to the $K$-band.  
The quoted spread is mainly caused by a difference in the 
zero-point, since the slopes are quite similar.   

The plotted intensity-weighted $RIJHK$ mean magnitudes can be
used to predict multi-band PL relations.
The evolutionary and pulsation parameters of the helium enhanced models will 
be provided in Marconi et al. (2018, in preparation) together with the 
bolometric mean magnitude, and the transformation into different optical, NIR 
and MIR photometric systems. Moreover, we plan to discuss the luminosity 
amplitudes for both canonical and He-enhanced models and their impact on the 
Bailey diagram. Finally, we plan to provide for the He-enhanced models the 
same distance diagnostics provided by \citep{marconi15}.  
Table \ref{plr} gives the coefficients of the 
global\footnote{This sample includes both fundamental and first 
overtone pulsators. The latter group was fundamentalized, i.e. 
$\log P_F = \log P_{FO}\;+\;0.127$} PL relations including both 
the metallicity and the helium terms 
($M_X$=$a$~+~$b \log P$~+~$c$[Fe/H]~+~$d \log$ Y, 
where X is the selected photometric band). 

A glance at the coefficients listed in this Table and to the models 
plotted in Fig.~\ref{FIG1} disclose three relevant features.

{\em i)} The dependence on the period becomes, as expected, systematically steeper 
when moving from optical to NIR bands. This means that NIR PL relations are 
intrinsically more accurate than optical ones. The reason is twofold: 
1) the standard deviation decreases by a factor of two when moving from 
the $R$/$I$ to the $H$/$K$ bands; 
2) the coefficient of the metallicity term in the NIR bands attains similar 
values. 

{\em ii)} The helium dependence decreases by roughly one dex when moving 
from the $R$- to the $K$-band. This means that an 
increase in helium content causes, at fixed period and 
metal content, an increase in the R-band of the order of a few tenths of a 
magnitude. The same increase causes a variation 
of the order of $\approx$0.05--0.08 mag 
in the NIR bands. Such an increase 
might introduce a mild systematic effect in distance 
determinations, but it appears negligible because it is similar to the 
standard deviations.

It has been suggested that the second stellar generation in GCs 
are made of materials that are enriched in helium, nitrogen and sodium 
and depleted in carbon and oxygen. The enhancement in helium can be
of the order of 0.05--0.10 \citep{renzini2015}. 
However, there are reasons to believe that 
the quoted dependence of NIR PL relations on helium can be considered 
as a solid upper limit on the RRL distance scale. The reasons are the 
following.   

{\em i)}--The second stellar generation appears to be ubiquitous in 
GCs, but the current spectroscopic evidence indicates that they are 
very rare in the Galactic field \citep[][]{gratton16}. This means 
that only a few percent of the galactic stellar content might be 
helium enhanced. 

{\em ii)}--He-enhanced stellar populations in the metal-poor and in 
the metal-intermediate regime cross the IS only during 
off-ZAHB evolution. This means that the evolutionary time spent inside 
the IS is at least one order of magnitude smaller compared
with the canonical ones (see \S~2).
Note that the 
He-enhanced (Y=0.30) ZAHB crosses the IS in the metal-intermediate 
and in the metal-rich regime. This means that the probability to produce 
He-enhanced RRLs in the metal-poor regime is quite limited.
Moreover, the crossing 
of the IS at brighter magnitudes (lower
surface gravities) causes a systematic shift in the period distribution of 
He-enhanced RRLs towards longer periods. This also means that He-enhanced stellar 
populations are more prone to produce type II Cepheids  (P $>$ 1 day) than 
RRLs.

\begin{deluxetable}{cccccc}
\tablecaption{
Coefficients of the predicted global (fundamental plus first overtone) 
Period-Luminosity-Metallicity-Helium (PLZY) relations for RRLs in the form 
$M_X$=$a$~+~$b \log P$~+~$c$[Fe/H]~+~$d \log$Y, 
where X is the selected band.   
\label{plr}}
\tablehead{
Band & \colhead{a} & \colhead{b} & \colhead{c} & \colhead{d} & \colhead{$\sigma$}\\
 }
\startdata
$R$ & -0.63$\pm$0.13 &-1.30$\pm$0.03 &0.195$\pm$0.007 & -1.34$\pm$0.07&0.13 \\
$I$ & -0.80$\pm$0.11 &-1.58$\pm$0.03 &0.190$\pm$0.005 & -1.10$\pm$0.06&0.11\\
$J$ &-1.00$\pm$0.08 &-1.92$\pm$0.02 &0.187$\pm$0.004 & -0.82$\pm$0.04&0.08\\
$H$ &-1.14$\pm$0.06 &-2.23$\pm$0.02 &0.188$\pm$0.003 & -0.55$\pm$0.03&0.06\\
$K$ &-1.16$\pm$0.06 &-2.26$\pm$0.02 &0.185$\pm$0.003 & -0.52$\pm$0.03&0.06\\
\enddata

\end{deluxetable}

\subsection{Mean magnitude---${M_B}, {M_V}$---metallicity relations}

The visual mean-magnitude metallicity (${M_V}-[Fe/H]$) relation was foreseen 
by \citet{baade58} and it was the most popular distance diagnostic for old 
stellar populations \citep{sandage90}, but it is also prone to a number of 
potential systematic errors~\citep{caputo00,bono03c,dicriscienzo04,marconi15}.

\begin{figure}
\includegraphics[scale=0.9]{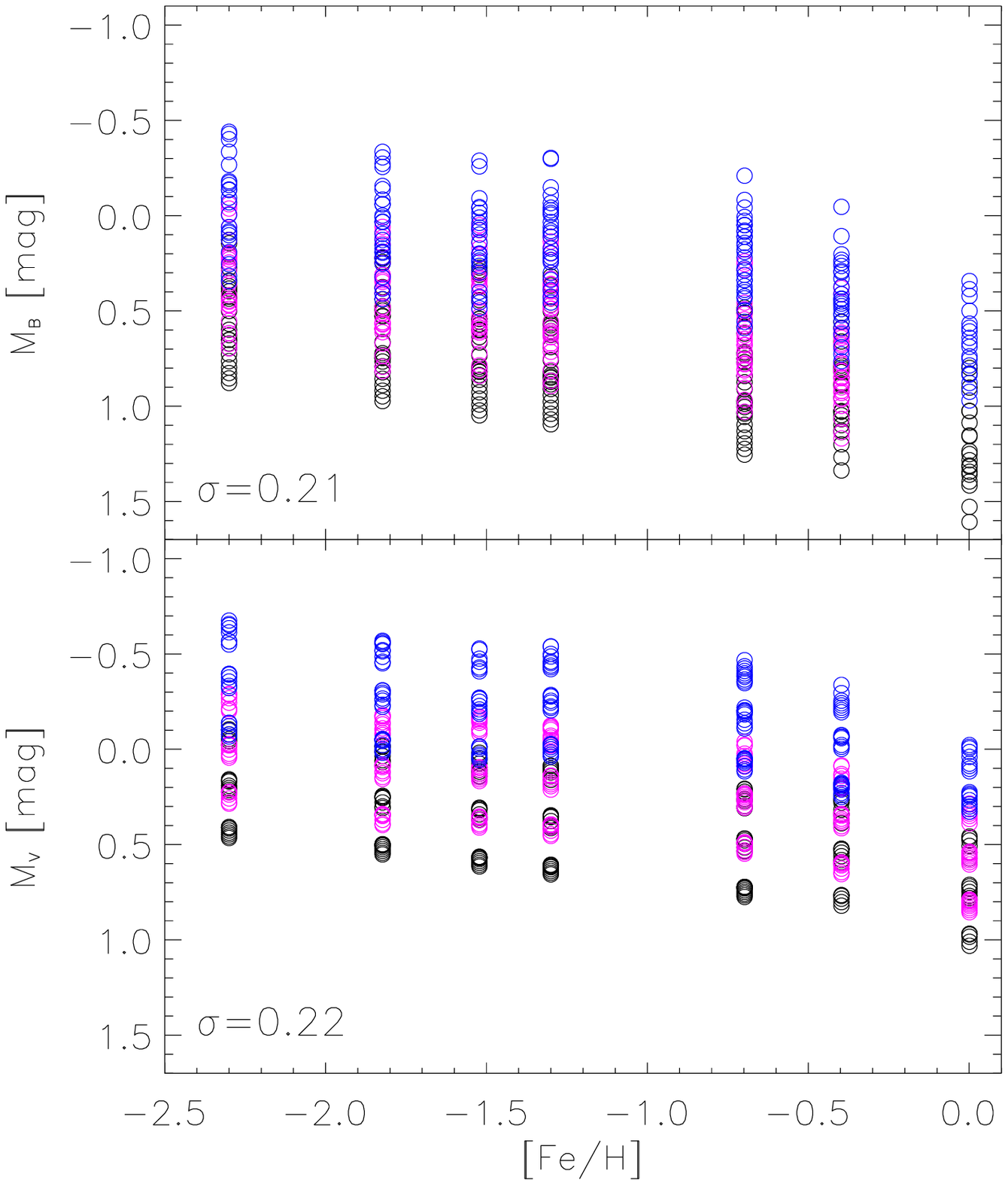}
\caption{Top: Predicted (fundamentals plus first overtones) B-band 
mean magnitude-metallicity (${M_B}-[Fe/H]$) relation. Symbols are 
the same as in Figure~\ref{FIG1}. The standard deviation is labeled 
in the bottom left corner. Bottom: Same as the top, but for the 
${M_V}-[Fe/H]$ relation.} \label{FIG3}
\end{figure}

We have already mentioned that an increase in He content causes, at fixed 
metallicity, an increase in stellar luminosity, and in turn, in the pulsation 
period. A glance at the predicted magnitudes plotted in the bottom panel of 
Figure~\ref{FIG3} shows the impact of the He content. Canonical and 
He-enhanced models, as expected, partially overlap due to off-ZAHB evolution. 
However, He-enhanced models with Y=0.30 (pink open circles) are on average 
$\sim$0.15 mag brighter than canonical ones, while those with Y=0.40 (blue 
open circles) are almost half magnitude brighter.  

The outcome is the same if we use the $B$-band, but it should be cautiously 
treated for a possible color dependency \citep{catelan04}.
Data plotted in the top panel of Figure~\ref{FIG3} display that He-enhanced 
models are 0.2 (Y=0.30) and 0.6 (Y=0.40) mag systematically brighter than 
canonical ones. To investigate on a more quantitative bases the dependence of the mean magnitude 
metallicity (MZ) relations, we derived new analytical relations including a 
He term (MZY), namely:

$$M_B(RR)=0.28[Fe/H]-2.91logY-0.61$$ with an r.m.s=0.21 mag

$$M_V(RR)=0.22[Fe/H]-2.94logY-1.08$$ with an r.m.s=0.22 mag

Data plotted in Figure~\ref{FIG3} and the above MZY relations 
disclose a few relevant predictions concerning the possible 
occurrence of He-enhanced RRLs. 

{\em i)}  Stellar systems hosting a sizable sample of RRLs covering 
a broad range in He abundance should show, at fixed metal content, 
a spread in visual and in $B$-band magnitudes that is on average a factor 
of two larger than canonical models. A detailed set of synthetic HB models is 
required to constrain the variation as function of both metal and He content.
However, empirical evidence dating back 
to \citep{sandage90} indicate that the spread in visual magnitude 
showed by cluster RRLs ranges from 0.2 mag in the metal-poor 
regime to 0.6 mag in the metal-intermediate regime. This trend 
was soundly confirmed by synthetic HB models by \citet{bono97a}

{\em ii)} Stellar systems hosting stellar populations with significantly different 
He contents should show multi-modal magnitude distributions inside the IS. 
Indeed, He-enhanced models are characterized by 
ZAHBs that are systematically brighter. This difference in magnitude 
cannot be mixed up with a difference in metallicity, since the evolutionary 
lifetime of He-enhanced models inside the IS is, at fixed metal 
content, systematically shorter than canonical ones.

\section{Final remarks and conclusions}

We have presented new sets of He-enhanced (Y=0.30, Y=0.40) nonlinear, 
time-dependent convective hydrodynamical models of RRLs 
covering the same range of metal abundances investigated by
\citet{marconi15}. The model mean magnitudes in the $RIJHK$ bands
were used to obtain new Period-Luminosity-Metallicity-Helium (PLZY) relations 
in these filters (see Table \ref{plr}). The main effect of an increase in 
He is an increase in the luminosity level, and in turn, in the
predicted pulsation period. Therefore, an increase in primordial He content from the canonical 
value (Y=0.245) to He-enhanced (Y=0.30,0.40) causes a minimal change in the 
coefficients of both period and metallicity terms, since the He--enhanced 
models obey similar PLZ relations. 
Owing to the sensitivity of the luminosity level to He 
variations, the classical relations connecting the $B$ and $V$ mean magnitudes 
to metallicity and the $R$-band PLZ relation display a significant He dependence.
The He-enhanced models 
models are, at fixed metallicity, 0.2$\div$0.5 mag brighter than canonical ones.

This is an interesting opportunity, because Gaia is going to 
provide accurate geometrical distances to calibrate both the zero-point 
and the slopes of the diagnostics adopted to estimate individual RRL distances. 
Spectroscopic RRL abundances based on ground-based measurements 
\citep{magurno2018} will pave the way for an empirical calibration of the 
PLZ relations. This means the opportunity to determine distance, reddening and 
chemical composition (metal, helium) for field RRLs for which are simultaneously 
available optical ($BVRI$) and NIR ($JHK$) mean magnitudes. Note that this approach 
applies to RRL in nearby stellar systems, and in turn, the opportunity to 
investigate the helium-to-metal enrichment ratio currently adopted in 
evolutionary and pulsation calculations is universal.


\bibliographystyle{aa}


 \end{document}